\documentclass[twocolumn]{aastex61}
\usepackage{graphicx}
\usepackage{epstopdf}
\usepackage{soul}
\usepackage{amsmath}

\received{????}
\revised{????}
\accepted{????}

\submitjournal{AJ}
\shorttitle{Chiron and its Rings}
\shortauthors{Wood et al.}

\begin{document}

\title{THE DYNAMICAL HISTORY OF 2060 CHIRON AND ITS PROPOSED RING SYSTEM}

\correspondingauthor{Jeremy Wood}
\email{jeremy.wood@kctcs.edu}

\author[0000-0003-1584-302X]{Jeremy Wood}
\affiliation{Hazard Community and Technical College, 1 Community College Drive Hazard, KY USA 41701}
\affiliation{University of Southern Queensland, Computational Engineering and Science Research Centre, West St, Toowoomba, QLD 4350, Australia}

\author{Jonti Horner}
\affiliation{University of Southern Queensland, Computational Engineering and Science Research Centre, West St, Toowoomba, QLD 4350, Australia}
\affiliation{Australian Centre for Astrobiology, UNSW Australia, Sydney, NSW 2052, Australia}

\author{Tobias C. Hinse}
\affiliation{Korea Astronomy and Space Science Institute, 776 Daedukdae-ro, Yuseong-gu, Daejeon 305-348, Republic of Korea}

\author{Stephen C. Marsden}
\affiliation{University of Southern Queensland, Computational Engineering and Science Research Centre, West St, Toowoomba, QLD 4350, Australia}

\keywords{minor planets, asteroids: individual: 2060 Chiron, planets and satellites: dynamical evolution and stability, planets and satellites: rings}


\newcommand{\change}[1]{\textcolor{red}{\textbf{#1}}}

\begin{abstract}
The surprising discovery of a ring system around the Centaur 10199 Chariklo in 2013 led to a reanalysis of archival stellar occultation data for the Centaur 2060 Chiron by \citet{2015A&A...576A..18O}. One possible interpretation of that data is that a system of rings exists around Chiron.\par
In this work, we study the dynamical history of the proposed Chiron ring system by integrating almost 36,000 clones of the Centaur backwards in time for 100 Myr under the influence of the Sun and the four giant planets. The severity of all close encounters between the clones and planets while the clones are in the Centaur region are recorded along with the mean time between close encounters.\par
We find that severe and extreme close encounters are very rare, making it possible that the Chiron ring system has remained intact since its injection into the Centaur region, which we find likely occurred within the past 8.5 Myr. Our simulations yield a backwards dynamical half-life for Chiron of 0.7 Myr. The dynamical classes of a sample of clones are found. It is found that, on average, the Centaur lifetimes of resonance hopping clones are twice those of random-walk clones because of resonance sticking in mean motion resonances. \par
In addition, we present MEGNO and chaotic lifetime maps of the region bound by 13 au $\le a \le$ 14 au and $e \le 0.5$. We confirm that the current mean orbital parameters of Chiron are located in a highly chaotic region of $a-e$ phase space.\par 
\end{abstract}

\section{INTRODUCTION}
The study of small bodies of the solar system was changed forever in 1977, with the discovery of a large icy object moving on an orbit between those of Saturn and Uranus \citep{1979IAUS...81..245K}. That object was subsequently named Chiron. It was soon realised that its orbit was dynamically unstable, with a mean half-life of 0.2 Myr, which is far shorter than the age of the solar system \citep[e.g.][]{1979AJ.....84..134O,1990Natur.348..132H}. For over a decade, 2060 Chiron was an oddity - but following the discovery of 5145 Pholus in 1992, a growing population of such objects in the outer solar system has been discovered - a population now known as the Centaurs. \par
Over the years, a number of different schemes have been proposed to define Centaurs \citep[e.g.][]{2003MNRAS.343.1057H,2005AJ....129.1117E,2007prpl.conf..895C,2008ssbn.book...43G}. Across all these schemes, it can be generally said that Centaurs have orbits between the giant planets Jupiter and Neptune. For this work, we follow the definition used by the Minor Planet Center, which considers objects to be Centaurs if they move on orbits with perihelia beyond the orbit of Jupiter and with semi-major axes within the orbit of Neptune\footnote{http://www.minorplanetcenter.net/iau/lists/Unusual.html (accessed 17th December 2016)}. Those objects in this region that are trapped in 1:1 resonance with one of the giant planets (the Trojans) are excluded from the list, and are not considered to be Centaurs. Using this definition, over 220 objects can presently be classified as Centaurs\footnote{http://www.minorplanetcenter.net/iau/lists/t\_centaurs.html (Accessed October 8, 2017)}. \par
The Centaurs move on highly chaotic orbits which are frequently perturbed by the gravitational influence of the four giant planets. The strongest perturbations typically occur as a result of close approaches between the Centaurs and those planets \citep[e.g.][]{1962ASPL....8..375M,2004MNRAS.355..321H}. The instability of the Centaur region is exemplified by the fact that Centaurs have dynamical lifetimes and half-lives much less than the age of the solar system, with values typically $\ll$100 Myr \citep{1996ASPC..107..233D,1997Icar..127...13L,2003AJ....126.3122T, 2004MNRAS.354..798H,2007Icar..190..224D,2009Icar..203..155B,2015A&A...583A..93P}. \par
It is therefore clear that these objects are ephemeral in nature, and that their ranks must be replenished over time from other sources. Proposed source populations for the Centaurs include the Oort Cloud \citep{2005MNRAS.361.1345E,2012MNRAS.420.3396B,2014Ap&SS.352..409D,2014Icar..231...99F}, the Jupiter Trojans \citep{2004MNRAS.354..798H,2006MNRAS.367L..20H,2012MNRAS.423.2587H}, the Neptune Trojans \citep{2010MNRAS.402...13H,2010MNRAS.405.1375L,2012MNRAS.422.2145H}, the Scattered Disk \citep{2007Icar..190..224D,2008ApJ...687..714V} and other populations in the Edgeworth-Kuiper Belt \citep{1997Icar..127...13L,2008ApJ...687..714V}. Of these many source regions, it is thought that the majority of Centaurs originate within the Scattered Disk \citep{2007Icar..190..224D,2008ApJ...687..714V}. \par
After these small bodies escape from one of the more stable source populations into the Centaur region, they will typically spend on the order of $\sim10^6$ years as a Centaur before diffusing out of that region \citep{2003AJ....126.3122T}. The final fates of Centaurs are varied - some will collide with the Sun or one of the planets, or will be torn apart by tidal forces during a planetary close encounter, whilst others will be thrown onto orbits beyond Neptune or be ejected from the solar system entirely \citep{1994PlR....14f...8N,2004MNRAS.354..798H,2008ApJ...687..714V,2016DPS....4812023W}. \par
During the course of their evolution, studies have shown that at least one-third of the Centaurs will evolve onto cometary orbits with perihelia in the inner solar system \citep{2004MNRAS.355..321H,2004come.book..659J,2009Icar..203..155B}. As such, the Centaurs are generally regarded as the principal parent population for the short period comets \citep{2003AJ....126.3122T,2004A&A...413.1163G,2004MNRAS.354..798H,2008ApJ...687..714V,2009Icar..203..155B,2009AJ....137.4296J,2011AstSR...7..230K}. \par
Indeed, several Centaurs (including Chiron) have been observed exhibiting cometary activity \citep[e.g.][]{2009AJ....137.4296J,2015MNRAS.454.3635S,2017AJ....153..230W}. 
%
Given the extreme dynamical instability exhibited by the Centaurs, coupled with the frequent close encounters they experience with the giant planets, the discovery in 2013 of a system of rings orbiting the Centaur 10199 Chariklo came as a huge surprise \citep{2014Natur.508...72B}. Those rings, revealed by unexpected dimmings of a star occulted by Chariklo prior to, and immediately after the occultation event, are narrow and dense, and lie at radii of $\sim$ 391 and $\sim$ 405 km.\par
It is still unknown whether the rings formed recently, or pre-date Chariklo's injection into the Centaur region, though rings have also recently been discovered around the dwarf planet Haumea \citep{2017Natur.550..219O} which orbits beyond Neptune. This suggests that rings around small bodies could form in the Trans-Neptunian region.\par
Furthermore, a recent dynamical study has shown that such rings could readily survive with Chariklo through its entire evolution in the Centaur region, since sufficiently close encounters to disrupt the rings are rare \citep{2017AJ....153..245W}.\par
\par
The chance discovery of Chariklo's ring system prompted a reanalysis of stellar occultation data obtained for 2060 Chiron in 1993, 1994 and 2011 by \citet{2015A&A...576A..18O}. The original analysis of that occultation data found dips in the light curve that, it was thought, corresponded to regions outside the nucleus which were then interpreted as comet-like dust jets \citep{1995Natur.373...46E,1996Icar..123..478B} or symmetrical jet-like features \citep{2015Icar..252..271R}. The recent reanalysis of this data suggests that it might also be interpreted as evidence for a ring system similar to that of Chariklo, with a mean radius of 324 $\pm$ 10 km \citep{2015A&A...576A..18O}.\par 
The origin of this proposed ring structure could be the result of a tidal disruption of Chiron due to a close encounter with a planet \citep{HyodoR:2016}, a collision between Chiron and another body \citep{2017A&A...602A..27M}, a collision between an orbiting satellite and another body \citep{2017A&A...602A..27M}, the tidal disruption of an orbiting satellite \citep{ElMoutamidM:2014} or debris ejected from Chiron itself due to cometary activity \citep{PanM:2016}.\par
Over time, rings can widen due to viscous spreading \citep{2017ApJ...837L..13M}. This process can occur on timescales as short as hundreds of years. However, the extent of the rings can be constrained, keeping them far more narrow, if shepherd satellites are present \citep{FrenchRG:2003, JacobsonRA:2004,ElMoutamidM:2014,2017ApJ...837L..13M}. At the present time no shepherd satellites are known to exist orbiting any Centaur and hence their possible dynamical role will not be considered in this study.\par 
Given the extreme dynamical instability exhibited by Chiron, it is interesting to consider whether its ring system could survive through the entirety of its life as a Centaur. If deep close encounters with the giant planets are sufficiently frequent, then it might be possible to place a constraint on the age of any rings around Chiron on the basis of its past dynamical history.\par
As a result, in this work, we follow \citet{2017AJ....153..245W}, and examine the dynamical history of Chiron and its proposed ring system. In doing so, we explore the likelihood that its rings could be 'primordial' (i.e. could date back to before the object was captured as a Centaur) barring ring dispersal by viscous spreading. Our results also allow us to explore the likely source population of Chiron, and to confirm its status as one of the most dynamically unstable Centaurs.\par
In section 2, we present the physical and orbital properties of 2060 Chiron. In section 3, we discuss the means by which we can measure the severity of close encounters between ringed small bodies and planets, and in section 4, we discuss the two dynamical classes that have been proposed for the Centaurs. We present our methodology in section 5, then present and discuss the results of our numerical integrations of Chiron in section 6. Finally, in section 7, we present our conclusions, and discuss possible directions for future work.

\section{THE PROPERTIES OF 2060 CHIRON}




\subsection{ORBITAL PROPERTIES}
After Chiron was discovered, pre-discovery images dating back as early as the late 19th century allowed its orbit to be well constrained {\citep{1977IAUC.3151....2L,1979IAUS...81..245K}. It was soon found that the orbit of Chiron was unlike the orbit of any known small body at the time. Its aphelion lay between Saturn and Uranus while its perihelion lay just interior to Saturn's orbit.\par
Since its discovery, more observations of Chiron have allowed its orbit to be even further refined. The current best-fit orbital properties of Chiron are shown in Table~\ref{chiron_orbital} and were taken from the Asteroids Dynamic site \citep{2012IAUJD...7P..18K}.\par
Using the semi-major axis, $a$, and eccentricity, $e$, from Table ~\ref{chiron_orbital}, the perihelion and aphelion distances are found to be 8.4 au and 18.86 au respectively. The semi-major axis is about 0.01 au away from the interior 5:3 mean motion resonance of Uranus located at about 13.66 au. 
The eccentricity of Chiron's orbit lies in the middle of the eccentricity range for the orbits of the known Centaurs, 0.01 - 0.73\footnote{http://www.minorplanetcenter.net/iau/lists/Centaurs.html (accessed 9 August, 2017)} and is high enough to cause Chiron to cross the orbits of both Saturn and Uranus. These giant-planet perturbations and close-approaches have a significant effect on the dynamical evolution of Chiron's orbit \citep{1979AJ.....84..134O,1979Icar...40..345S,2002EM&P...90..489K} which is reflected in the relatively short dynamical lifetime of $\sim$ 1 Myr \citep{1990Natur.348..132H,2004MNRAS.354..798H}.\par
Furthermore, the half-life of its orbit is 1.03 Myr in the forward direction and 1.07 Myr in the backward direction \citep{2004MNRAS.354..798H}. Both times are much less than the age of the solar system.\par
The instability of Chiron's current orbit makes it highly unlikely that its orbit is primordial. Instead, the general consensus is that Chiron follows a chaotic orbit and originated in the Kuiper Belt \citep{1979AJ.....84..134O,1990Natur.348..132H,1996P&SS...44.1547L,2001P&SS...49.1325S,2002Icar..160...44D,2002EM&P...90..489K}. \par
Using the taxonomy of \citet{2003MNRAS.343.1057H}, Chiron is classified as an object in the SU$_{\textnormal{IV}}$ class. This means that its dynamics are controlled by Saturn at perihelion and by Uranus at aphelion. 
The subscript IV means that the Tisserand parameter with respect to Saturn is $>$ 2.8 \citep{2003MNRAS.343.1057H}. The Tisserand parameter, $T_{p}$, is a quantity calculated from the orbital parameters of a small body and those of a planet it could encounter. It is defined by:

\begin{equation}
T_p = \frac{a_p}{a} + 2\textnormal{cos}(i-i_p) \sqrt{\frac{a}{a_p}(1-e^2)}
\end{equation}

\noindent{\citep[e.g.][]{MurrayCD:1999}}. Here, $a_p$ is the semi-major axis of a planetary orbit, $i$ the inclination of the small body orbit and $i_p$ the inclination of the planetary orbit.\par
To first order, the Tisserand parameter of an orbit with respect to a given planet is expected to be conserved through an encounter with that planet, with the precise value giving an indication of the maximum strength of encounters that are possible with that planet.\par
Broadly, if $T_{p} > 3$, then particularly close encounters are not possible between the two objects, whilst for $2.8 \le T_{p} \le 3$, then extremely close encounters can occur that might lead to the object being ejected from the solar system in a single pass \citep{2003MNRAS.343.1057H}.
\begin{table}
\caption{The orbital elements of Chiron for epoch 2457600.5 JD, based on an observational arc length of 44,305.9 days taken from the Asteroids Dynamic site (accessed 31st December, 2015). Here, $a$ is the semi-major axis, $e$ the eccentricity and $i$ the inclination of the orbit. $\Omega$, $\omega$ and $M$ are the longitude of ascending node, argument of perihelion and Mean anomaly respectively. Each uncertainty is the standard deviation around the best-fit solution.}\label{chiron_orbital}
\begin{tabular} {|c|c|c|c|}
\hline
Property&Value&Units\\
\hline
$a$&13.639500 $\pm$ $(1.48\times 10^{-6})$&au\\
$e$&0.38272700 $\pm$ $(9.62\times 10^{-8})$&\\
$i$&6.947000 $\pm$ $(6.67\times 10^{-6})$&deg\\
$\Omega$&209.21600 $\pm$ $(6.05\times 10^{-5})$&deg\\
$\omega$&339.53700 $\pm$ $(6.19\times 10^{-5})$&deg\\
$M$&145.97800 $\pm$ $(2.97\times 10^{-5})$&deg\\
\hline
\end{tabular}
\label{chiron_orbital}
\end{table}

\subsection{DENSITY, SIZE AND MASS}
Unlike the relatively high precision with which the orbital parameters of Chiron are known, the physical properties remain much more poorly constrained. The diameter of Chiron has had to be estimated based on an assumed albedo. Though a strong effort to determine the size of Chiron has been made over the past two decades, efforts have been hampered by the interference from possible material located outside the nucleus, cometary activity and Chiron's elongated shape \citep{2013A&A...555A..15F,2015A&A...576A..18O}.\par
Radius measurements ranging from 71 km \citep{2004A&A...413.1163G} to a constraint of $<$ 186 km \citep{1991Sci...251..777S} have been reported. \citet{2015A&A...576A..18O} report an overall average effective spherical radius of 90 km which we adopt for this work. \par
Because of the large uncertainty in the size and mass of Chiron, Chiron's overall density is also poorly known. \citet{1997AJ....113..844M} in their study of a coma around Chiron report a bulk density in the range 500 - 1,000 kgm$^{-3}$. Using a spherical radius of 90 km this corresponds to a mass range of 1.53$\times 10^{18}$ kg - 3.05$\times 10^{18}$ kg.\par

\section{Measuring the Severity of Close Encounters with Planets}

Currently, it is unknown what role, if any, the sporadic activity of Chiron played in the formation of any ring structure around the body.  Rings could have formed either before or after Chiron entered the Centaur region. But given that Chiron presently lies in a chaotic and unstable orbit prone to planetary close encounters, it is of interest to determine the likelihood that such encounters could severely damage or destroy any orbiting ring structure.\par
To accomplish this, a method to gauge the severity of such an encounter is needed. Primarily, the severity of a close encounter between a ringed small body and a planet is determined by the minimum approach distance between the small body and planet, $d_{min}$.\par
If the small body is in a parabolic or hyperbolic orbit relative to the planet (it hasn't been captured as a satellite), then the velocity at infinity of the small body relative to the planet also plays a role in determining the encounter severity, albeit to a lesser extent than the depth of the encounter.\par 
\citet{2017AJ....153..245W} ignored velocity effects and developed a severity scale based on $d_{min}$ relative to the Hill radius, $R_H$, tidal disruption distance, $R_{td}$, the ring limit, $R=10R_{td}$, and Roche limit, $R_{roche}$. This scale is shown in Table~\ref{CE_severity}.
\begin{table} [h]
\caption{{A scale ranking the severity of a close encounter between a ringed small body and a planet based on the minimum distance obtained between the small body and the planet, $d_{min}$, during the close encounter. $R_H,R=10R_{td},R_{td}$ and $R_{roche}$ are the Hill radius of the planet with respect to the Sun, ring limit, tidal disruption distance and Roche limit respectively.}}\label{CE_severity}
\begin{tabular} {|c|c|}
\hline
$d_{min}$ Range&Severity\\
\hline
$d_{min} \ge R_H$&Very Low\\
$R\le d_{min}< R_H$&Low\\
$R_{td}\le d_{min}< R$&Moderate\\
$R_{roche}\le d_{min}< R_{td}$&Severe\\
$d_{min}<R_{roche}$&Extreme\\
\hline
\end{tabular}
\end{table}

The Hill radius defines a sphere of influence centered on a secondary body of mass $m_{s}$ in an orbit with orbital radius $R_{radial}$ around a primary body of mass $M_p$ in the planar problem. The Hill radius is approximately given by:
\begin{equation}
\Large{R_{H}\approx R_{radial}(\frac{m_{s}}{{3M_p}})^{\frac{1}{3}}}
\label{hilleqn}
\end{equation}
\noindent{\citep[e.g.][]{MurrayCD:1999}}. For non-circular orbits, $R_{radial}$ is approximated using the semi-major axis of the orbit. Loosely defined, the Hill radius is the distance around a secondary body (relative to a primary body) within which satellites can orbit without their orbits being completely disrupted by tidal forces due to the primary body. In the case where the secondary body is a planet and the primary body the Sun, it is found that all known planetary satellites follow this rule, being contained well within the Hill sphere's of their host planets. For other objects moving in the system, the Hill radius of a planet can be used to indicate the region of space around its orbit into which other objects move at their peril.\par
Typically, encounters at a distance greater than $\sim$ 3 Hill radii will have only a limited effect on the long term stability of an object, whilst orbits that approach within this distance are typically dynamically unstable, unless close approaches are prevented by mutual mean-motion resonances between the objects concerned \citep[e.g.][]{1971AJ.....76..167W,1995AJ....110..420M,HUAqr,2012ApJ...754...50R,2012ApJ...761..165W}.\par
The ring limit is a relatively new critical distance introduced by \citet{AraujoRAN:2016} and used by \citet{2017AJ....153..245W} to examine the stability of Chariklo's ring system against close encounters. It is loosely defined as lying at ten tidal disruption distances from a given planet, and represents an upper limit on the minimum approach distance for close encounters for which the effect on a ring of a minor body is just noticeable (meaning the maximum change in orbital eccentricity of the orbit of any ring particle = 0.01). Here, we apply the ring limit to study the influence of close encounters between Chiron and the giant planets.\par
Given a typical solar system small body, the tidal disruption distance, $R_{td}$, lies well within the Hill radius for a given planet. When the separation between a small body and a planet is closer than $R_{td}$, a secondary body-satellite binary pair of total mass $m_{s}+m_{sat}$ and semi-major axis $a_B$ can be permanently disrupted by tidal forces in one pass. It should be noted, in passing, that, defined in this manner, the ring limit and tidal disruption distances have no meaning for close encounters between planets and small bodies with no rings or satellites.\par
$R_{td}$ can be approximated as the secondary-primary body separation at which a satellite orbiting the secondary body would lie at the outer edge of the secondary body's Hill sphere. $R_{radial}$ in Equation~\ref{hilleqn} is then by definition $R_{td}$, and $R_{H}$ is approximated by $a_B$. Solving for $R_{td}$ yields:

\begin{equation}
\Large{R_{td}\approx a_B(\frac{3M_p}{m_{s}+m_{sat}})^{\frac{1}{3}}}\label{tidal_disrupteqn}
\end{equation}

\noindent{\citep[e.g.][]{PhilpottCM:2010}}. Closer still to the primary body, the Roche limit is the distance from the primary within which a secondary body held together only by gravity would be torn apart by tidal forces. For a rigid secondary body, the equation for the Roche limit with respect to a primary body is approximately:

\begin{equation}
\Large{R_{roche} = 2.44R_{p}(\frac{\rho_p}{\rho_{s}})^{\frac{1}{3}}}\label{rocheeqn}
\end{equation}

\noindent{\citep{1849Roche...1..243,MurrayCD:1999}}. Here, $R_p$ is the physical radius of the primary body, $\rho_p$ the density of the primary body and $\rho_s$ the density of the secondary body.\par
Now that a severity scale for close encounters has been established, it can be used to study simulated close encounters between ringed Centaurs and the giant planets.

\section{The Two Dynamical Classes of Centaurs}

Throughout its lifetime as a Centaur, the frequency and severity of close encounters between Chiron and the giant planets will affect the stability of any ring structure around Chiron. The frequency of close encounters can be affected by a Centaur's so called dynamical class.\par
Previously it was shown that small bodies including Centaurs can be classified based on their perihelion, aphelion and Tisserand parameter \citep[as detailed in][]{2003MNRAS.343.1057H}.\par
However, as \citet{2009Icar..203..155B} showed, Centaurs may also be classified into one of two classes based on their long-term dynamical behavior. The first type consists of those Centaurs that randomly wander from orbit to orbit. The semi-major axes of these Centaurs' orbits increase and decrease in time with no particular pattern. These Centaurs are known as random-walk Centaurs.\par
Centaurs of the other type spend most of their time temporarily trapped in mean motion resonances of the giant planets and typically jump from one resonance to the other. A small body is in a mean motion orbital resonance with a planet if the ratio of the orbital period of the planet to the orbital period of the small body equals a ratio of two small integers \citep{MurrayCD:1999}.\par
Becoming temporarily trapped in a resonance is a behavior known as resonance sticking \citep{2007Icar..192..238L}. While trapped in a resonance, the semi-major axes of these Centaurs' orbits oscillate about a constant value which corresponds to the resonance location. These Centaurs are known as resonance hopping Centaurs. Since it is possible that resonance sticking can protect small bodies from close encounters with planets \citep{1995AJ....110..420M}, the dynamical class of a Centaur can have consequences for any ring structure around it.\par
The two types can also be more rigorously defined mathematically. As the semi-major axes of random-walk Centaurs wander aimlessly and those of resonance hopping Centaurs remain more constant, we would expect that on average the standard deviation of semi-major axis values of random-walk Centaurs would increase in time more predictably than those of resonance hopping Centaurs.\par
Mean standard deviation then, can be used as a tool to distinguish between the two dynamical types. Random-walk Centaurs are those Centaurs whose mean square standard deviation of semi-major axis, $\langle \sigma^2 \rangle$, varies as a power law in time. It is said that these Centaurs display generalized diffusion. This can be expressed mathematically as:

\begin{equation} 
\Large{\langle \sigma^2 \rangle = Dt^{2H}}
\label{gen_diff}
\end{equation}

Here, $t$ is time, $D$ is the generalized diffusion coefficient and $H$ is the Hurst exponent with $0 < H < 1$. Random-walk Centaurs can then be generally defined as those Centaurs for which the semi-major axis behavior is well described by generalized diffusion. Conversely it then goes that the behavior of the semi-major axis of resonance hopping Centaurs is not well described by generalized diffusion.\par
Centaurs of both types may also display both random walking and resonance sticking during their lifetime. To determine if a Centaur is in fact trapped in a particular mean motion resonance, care must be taken.\par
Resonances do not exist at a single point but have widths in phase space. For example, for any particular resonance, a Centaur can be trapped in the resonance over a range of semi-major axis values.\par
To positively determine if a small body is trapped in a resonance, two behaviors must be displayed. First, the semi-major axis of the small body orbit must oscillate about the resonance location, and second, the primary resonance angle must librate in time \citep{2013Icar..222..220S}.\par
The primary resonance angle is defined by $p\lambda -q\lambda_p - (p-q)\bar{\omega}$ where $p$ and $q$ are integers, $\lambda_p$ is the mean longitude of the planet's orbit, $\lambda$ is the mean longitude of the small body's orbit, and $\bar{\omega}$ is the longitude of perihelion of the small body's orbit \citep{MurrayCD:1999,2002MNRAS.335..417R,2009Icar..203..155B,2013Icar..222..220S}.\par
This angle is related to the perturbation of the orbit of a small body around a central body (like the Sun) by a third body (like a planet) in the planar 3-body problem. The reader is referred to \citet{MurrayCD:1999} for details.

\section{METHOD}


To study the dynamical history of Chiron and its ring system, a suite of numerical integrations were performed using the $n$-body dynamics package {\sc Mercury} \citep{ChambersJE:1999}.\par
35,937 massless clones of Chiron were integrated backwards in time for 100 Myr in the six-body problem (Sun, four giant planets, and clone). The integration time is justified as it is at least 100 times longer than the approximate half-life of Chiron \citep{1990Natur.348..132H,2004MNRAS.354..798H}.\par
The orbital elements of the individual clones were chosen from a range of three standard deviations below to three standard deviations above the accepted value of each orbital parameter of Chiron for epoch 2457600.0 JD taken from the Asteroid Dynamic Site \citep{2012IAUJD...7P..18K}.\par
To create our cloud of clones for Chiron, we varied each of the orbital elements, as follows. First, we sampled the $\pm 3 \sigma$ uncertainty range in semi-major axis, $a$. We tested eleven unique values of semi-major axis, ranging from $a - 3 \sigma$ to $a + 3 \sigma$, in even steps. At each of these unique semi-major axes, we tested eleven orbital eccentricities, which were again evenly distributed across the $\pm 3 \sigma$ uncertainty in that variable. At each of these $121 a-e$ pairs, we tested eleven unique inclinations also evenly spaced in the range $\pm 3 \sigma$. This gave a grand total of 1331 potential $a-e-i$ combinations for Chiron. At each of these values, we tested 27 unique combinations of $\Omega$, $\omega$ and $M$, creating a 3$\times$3$\times$3 grid in these three elements. The three values chosen for each of these three variables were the best-fit solution, and the two values separated by $3 \sigma$ from that value. In total, this gave us a sample of 35,937 unique orbital solutions for Chiron.\par
The time step was chosen to be 40 days which is approximately one-hundredth an orbital period of Jupiter - the innermost planet included in this study. Similar time steps have been used before in integrations of both Centaurs and Main Belt asteroids \citep{2000Icar..146..240T,2003AJ....126.3122T}.\par
Clones were removed from the simulation upon colliding with a planet, colliding with the Sun, achieving an orbital eccentricity $\ge 1$, or reaching a barycentric distance $>$ 1,000 au.\par
The masses and initial orbital elements of the four giant planets were found using the NASA JPL HORIZON ephemeris\footnote{http://ssd.jpl.nasa.gov/horizons.cgi?s\_body=1$\#$top (accessed 31st December 2015)} for epoch 2451544.5 JD. Inclinations and longitudes for both Chiron and the planets were relative to the ecliptic plane.\par
In order to set their starting orbital parameters for the simulation, the planets were integrated (within the heliocentric frame) to the epoch 2457600.0 JD - the epoch of the Chiron clones using the \textit{Hybrid} integrator within the \textsc{Mercury} $n$-body dynamics package \citep{ChambersJE:1999}. The accuracy parameter was set to 1.d-12, and the hybrid handover radius was set to three Hill radii.


Statistics on the close encounters were then taken by small body population of the solar system membership of the clone at the time of the encounter and by encounter severity. 
The different small body populations of the solar system used are defined in Table~\ref{ss_populations}.

\begin{table} [h]
\caption{Some different small body populations of the solar system. Here, $a$ is the semi-major axis of the clone during the close encounter. The semi-major axis and other orbital values of the clone's orbit just before the close encounter were not recorded. $a_J$ and $a_N$ are the semi-major axis of Jupiter and Neptune respectively; and $q$ is the perihelion distance of the clone. Inner SS means inner solar system, SP Comet means short period comet, TNO means Trans-Neptunian Object and Ejection means the clone was being ejected from the solar system at the time of the encounter.}
\begin{tabular} {|c|c|}
\hline
Name&Definition\\
\hline
Inner SS&$a\le a_J$\\
SP Comet&$a>a_J$ and $q<a_J$\\
Centaur&$a_J<a<a_N$ and $q>a_J$\\
TNO&$a\ge a_N$\\
Ejection&$e\ge 1$\\
\hline
\end{tabular}
\label{ss_populations}
\end{table}

Physical properties of the planets were taken from NASA\footnote{$https://ssd.jpl.nasa.gov/?planet\_phys\_par$ (accessed June 16, 2017)}. The mass of the Sun was also taken from NASA\footnote{$https://nssdc.gsfc.nasa.gov/planetary/factsheet/sunfact.html$ (accessed June 17 2017)}. For Chiron we selected a bulk density of 1,000 kgm$^{-3}$, which along with our selected radius of 90 km yielded a mass of 3.05$\times 10^{18}$ kg. This mass was used in equation \ref{tidal_disrupteqn} to determine the tidal disruption distance between Chiron and each planet. The density was used in equation \ref{rocheeqn} to determine the Roche Limit between Chiron and each planet.

\subsection{Determining the Half-Life and Origin of Chiron}
To determine the likely origin of Chiron, the chronologically earliest close encounter with a giant planet was analyzed for each clone, and the small body population of which the clone was a member at the time of the close encounter was found using the orbital parameters of the clone's orbit at the time of the encounter.\par
This then allowed the fraction of injection events from the various small body populations shown in Table~\ref{ss_populations} to be determined (in other words, it allowed us to determine the likely source population of Chiron). 

Note that Trojans could overlap with the Centaur small body population the way we have defined it. However, in order to have a close encounter, a small body must have already exited the Trojan region. \par
Furthermore, though the Jupiter and Neptune Trojans are possible feeder populations to the Centaurs \citep[e.g.][]{2006MNRAS.367L..20H,2010MNRAS.402...13H}, our study is unable to yield any information on the likelihood as either of these being the source of Chiron. Therefore Trojans were omitted as separate populations in Table~\ref{ss_populations}.\par
To determine the half-life of Chiron against removal from the simulation moving backwards in time, the number of clones remaining at a time $t$ was recorded as a function of time throughout the entire integration. Given $N_o$ as the initial number of clones at a time $t = 0$, the half-life can be determined by fitting the data to the standard radioactive decay equation:

\begin{equation}
\Large{N=N_oe^{ \frac{-0.693}{\tau}t }}
\label{half_life_eqn}
\end{equation}

\noindent{where $\tau$ is the half-life}. The time interval over which the decay of clones was exponential was obtained by the fit of the data to equation~\ref{half_life_eqn}. Then the fit was used to calculate the half-life.\par
Once the half-life was determined, it was used in equation~\ref{half_life_eqn} to determine the time at which 99.99\% of clones would be removed from the simulation assuming a constant half-life. This time was then set as the upper limit on the time at which Chiron entered the Centaur region.\par


\subsection{Finding the Dynamical Class}
A separate set of integrations was made using the IAS15 integrator in the \textsc{Rebound} $n$-body simulation package \citep{2012A&A...537A.128R,2015MNRAS.446.1424R} using the orbital values from a set of 1,246 Chiron clones from the previous integrations. \par
Three different samples of clones of $\sim$400 clones each were used - the first sample was taken from the first 1,000 clones, the second from the middle 1,000 clones and the third from the last 1,000 clones in the entire data set. The middle sample included the currently accepted orbital values of Chiron.\par
It was not necessary to find the dynamical class of every clone since the objective of these integrations is to compare and contrast the two dynamical classes and to explore specific examples of the behavior of clones in each class. Just a sampling of clones is sufficient for these purposes.\par
The output time was set to 300 years, and the time step to 0.1 year. In these integrations, clones were removed from the simulation upon colliding with the Sun, colliding with a planet, achieving an eccentricity $\ge 1$ or by leaving the Centaur region. Any clone which did not remain in the Centaur region for at least 100,000 years was not used. The dynamical class of each remaining clone was found using the method of \citet{2009Icar..203..155B}:

\begin{enumerate}
\item{Determine the time at which the clone was injected into the Centaur region, $T_{Centaur}$. Determine the number of data points in the time interval [0, $T_{Centaur}$].}
\item{Create a logarithmic interval of data points using [log(10), log(Data Points)].}
\item{Divide the interval into 16 equal logarithmic increments. Call the length of one of these increments $j_s$.}
\item{Create a window length of ten data points in units of time. Set this equal to the smallest window length.}
\item{Create each $z_{th}$ additional window length in units of data points, $w(z)_{datapts}$, by converting a logarithmic window into a window of data points using $w(z)_{datapts}=10^{1+z(j_s)}+1$ where $z\ge 1$.}
\item{Convert each window length from units of data points into units of time using $w(z)_{time}=w(z)_{datapts} \times $(output time). The interval each window covers is closed on one end and open on the other. For example, the first window time interval would be [0, $w(z)_{time})$.}
\item{Discard any window lengths more than 25\% of the data set.}
\item{Using the smallest window length, partition the time interval [0, $T_{Centaur}$] into equal windows of time and allow each window to overlap adjacent windows by half a window length.}
\item{Within each window determine the standard deviation, $\sigma$, of the semi-major axis, $a$.}
\item{Calculate the mean standard deviation, $\bar{\sigma}$, over all windows.}
\item{Repeat the process for all the window lengths.}
\item{Perform a linear regression on log($\bar{\sigma}$) vs. log($w(z)_{time}$).}
\item{The slope obtained from this regression is an approximation of the Hurst exponent.}
\item{A residual is the difference between an actual value and its expected value from the best-fit line. In this case, a residual of a particular value of log($\bar{\sigma}$) is found by finding the absolute value of the vertical distance from a value of log($\bar{\sigma}$) from the best-fit line. A Centaur is classified as being resonance hopping if the maximum value of any one residual is $\ge$ 0.08. Otherwise, the Centaur is classified as random-walk. This method is based on the results of \citet{2009Icar..203..155B}, and the reader is referred to that work for more details.}
\end{enumerate}

Selected resonance hopping clones were studied in more detail by examining intervals of time in which the semi-major axis oscillated about a nearly constant value.\par
The semi-major axis values for these intervals of time were then smoothed using the technique of \citet{2010MNRAS.404..837H} to determine if the clone was trapped in a mean motion resonance of a giant planet. The method is as follows:\par


\begin{enumerate}
\item{Qualitatively inspect graphs of semi-major axis vs. time for resonance hopping Centaurs and identify intervals of time, $\Delta T_{res}$, in which the semi-major axis seems to oscillate about a nearly constant value.}
\item{Select one of these intervals of time for study. Create a set of all semi-major axis data points during this time interval.}
\item{Initially, set the smoothed data set equal to the original data set.}
\item{By inspection, decide on a time window in units of data points. Set the window length to an odd number of data points and call this $w_N$.}
\item{Apply the window to the original data set at the first data point.}
\item{Evaluate the mean value of the semi-major axis over all data points within the window.}
\item{Set the value of the middle data point in this window (the $j_{(w_N-1) \times 0.5}$ data point) in the smoothed data set to this mean value.}
\item{Slide the window ahead by one data point in the original data set and set the value of the middle data point in this window in the smoothed data set equal to the mean semi-major axis over the entire window in the original data set.}
\item{Continue this process until the window ends on the last data point. If $j_{last}$ is the last data point then in the smoothed data set the $j_{last} - j_{(w_N-1) \times 0.5}$ data point is set to the mean value of the semi-major axis in the window in the original data set. Any data points before the $j_{(w_N-1) \times 0.5}$ data point and after the $j_{last} - j_{(w_N-1) \times 0.5}$ data point in the smoothed data set remain unchanged.}
\item{Try various window lengths until the smoothed data is as close to a cosine or sine wave in time as can be obtained by inspection.}
\item{Set the nominal location of the mean motion resonance equal to the mean value of the semi-major axis over the time interval $\Delta T_{res}$ in the smoothed data set.}
\item{Compare this location to known locations of mean motion resonances of the giant planets for identification. If the mean value is within 0.1 au of a resonance location then consider that resonance as a possible candidate.}
\item{Examine the primary resonance angle associated with each candidate resonance for librating behavior over the time interval. If the angle librates then consider the clone to be trapped in the resonance over the time of libration.} 
\end{enumerate}

The locations of mean motion resonances of the giant planets, $a_{res}$, were found using:

\begin{equation}
\Large{a_{res} = (\frac{j_1}{j_2})^{\frac{2}{3} }a_p}
\end{equation}\label{mmr_loc}

\noindent{\citep{MurrayCD:1999}}. Here, $a_p$ is the semi-major axis of a planet; and $j_1$ and $j_2$ are integers. In this work $j_1$ and $j_2$ were limited to values between 1 and 20. 

\subsection{MEGNO and Lifetime Maps}
The chaoticity and chaotic lifetime of Chiron's orbital evolution were studied by means of calculating global MEGNO and lifetime maps over a given parameter region. The MEGNO (Mean Exponential Growth of Nearby Orbits) \citep{2000A&AS..147..205C,2001A&A...378..569G,2003PhyD..182..151C,2004A&A...423..745G,2010MNRAS.404..837H} factor is a quantitative measure of the degree of chaos and has found wide-spread applications within problems of dynamical astronomy. The time averaged MEGNO parameter, $\langle Y \rangle$, is related to the maximum Lyapunov Characteristic Exponent, $\gamma$, by:

\begin{equation}
\Large{\langle Y \rangle = t\frac{\gamma}{2}}
\end{equation}

\noindent{as $t \rightarrow \infty$}. For more on Lyapunov characteristic exponents, we direct the interested reader to \citet{1995Icar..115..347W}. \par
The detection of chaotic dynamics is always limited to the integration time period. Quasi-periodic or regular motion could in principle develop into chaotic motion over longer time scales. The calculation of $\langle Y \rangle$ involves the numerical solution of the associated variational equations of motion.\par
Following the definition of MEGNO, the quantity $\langle Y \rangle$ asymptotically approaches 2.0 for $t \rightarrow \infty$ if the orbit is quasi-periodic. For chaotic orbits, $\langle Y \rangle$ rapidly diverges far from 2.0. In practice, the limit $t\rightarrow \infty$ is not feasible and $\langle Y \rangle$ is only computed up to the integration time (eventually ended by some termination criterion such as the event of an escape or collision).\par
A MEGNO map is created using the technique of numerical integration of a number of massless test particles starting on initial orbits which cover a rectangular grid in $a-e$ space, with other orbital parameters held constant. In this work, the Gragg-Bulirsh-St$\ddot{o}$er \citep{HairerE:1993} method was used to integrate 300,000 test particles for 1 Myr in the region of $a-e$ space bound by $13$ au $\le a \le 14$ au and $0 \le e \le 0.5$. The other orbital parameters were set to those of Chiron.\par
The resolution of the map was $600 \times 500$ ($a-e$). One test particle was integrated for each $a-e$ pair for a total of 300,000 $a-e$ pairs.\par
The time step varied and was determined using a relative and absolute tolerance parameter both of which were set to be close to the machine precision. A test particle was removed from the simulation if it collided with a planet or the Sun, was ejected from the solar system, or if $\langle Y \rangle > 12$ (indicating a strong degree of chaos).\par
When a test particle was removed, the time of removal and the $\langle Y \rangle$ value were recorded. If a test particle survived the entire simulation then its removal time was recorded as 1 Myr. We will call the removal time the ``chaotic lifetime'' which is not the same as dynamical lifetime. However, it can be said that the dynamical lifetime is equal to or greater than the chaotic lifetime.\par
A chaotic lifetime map was then generated in conjunction with the MEGNO map by color coding the lifetimes in the same $a-e$ grid used to create the MEGNO map. In the lifetime map the shortest removal times were color coded black and the longest yellow. The resulting lifetime and MEGNO maps can be seen in Figure~\ref{lifetime_map} and Figure~\ref{megno_map} respectively.

\section{RESULTS}

\subsection{Half-Life and Origin of Chiron}
The percentage of first close encounters by clone small body population membership is shown in Table~\ref{chiron_origin}. The TNO population has the highest percentage of first close encounters making it the most likely source population of Chiron.\par
34\% of clones were in a hyperbolic or parabolic orbit during their first close encounter, which indicates a potential origin within the Oort cloud. The Centaur and Inner solar system populations combined contributed just 3\% of the first close encounters. \par
The short period comet population claims 2\% of first close encounters. These three populations combined likely illustrate potential final destinations for Chiron in the future, since dynamical evolution that takes no account of the influence of non-gravitational forces is entirely time-reversible.

\begin{table} [h]

\caption{The percentage of first close encounters by clone small body population membership. The TNO population has the highest percentage of first close encounters making it the most likely source population of Chiron.}
\begin{tabular} {|c|c|c|}

\hline
Region&\% CE\\
\hline
Inner SS&1\\
SP Comet&2\\
Centaur&2\\
TNO&60\\
Ejection &34\\
\hline
\end{tabular}

\label{chiron_origin}
\end{table}

Figure~\ref{half_life} shows the natural log of the fraction of remaining clones vs. time over the last 2.5 Myr. The decay is exponential for the time interval [0.12 Myr, 0.5 Myr].
By 1 Myr ago, the decay curve departs markedly from this initial exponential decay. \par
This is typical and results from clones which have evolved onto more stable orbits. Because of this, these clones are no longer sampling the original phase space at the start of the decay.\par
To maximize the fit, the half-life during the exponential decay was determined on the interval [0.12 Myr, 0.367 Myr] and found to be about 0.7 Myr. Other larger intervals were tried and yielded the same result. This value is comparable to, but slightly shorter than, the value of 1.07 Myr reported by \citet{2004MNRAS.355..321H} for this quantity.\par 
Our smaller value is not surprising because \citet{2004MNRAS.355..321H} found their half-life using the longer time interval of 3 Myr which included a longer tail over which the half-life was markedly different from its initial value.

\begin{figure*} [h]
\begin{center}
\includegraphics[]{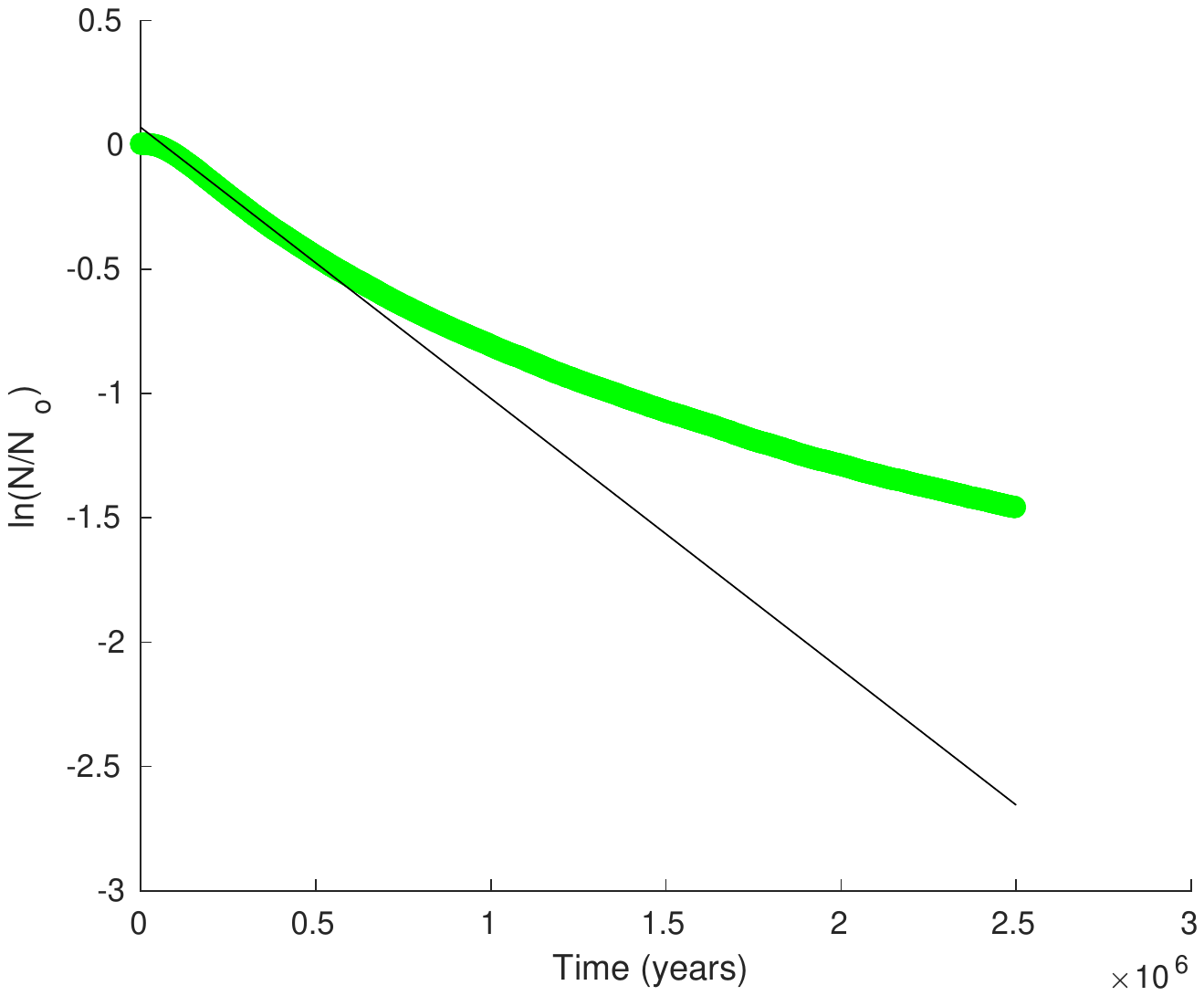}
\caption{The natural log of the fraction of remaining clones vs. time over the last 2.5 Myr. The decay is exponential through the interval [0.12 Myr, 0.50 Myr]. The half-life during the interval [0.12 Myr, 0.367 Myr] was found to be about 0.7 Myr. The solid line is the best-fit line for this interval and fits the data with a linear regression coefficient of 0.9999. By 1 Myr ago it can be seen that the decay is no longer exponential.}
\end{center}
\label{half_life}
\end{figure*}

786 clones, just 2$\%$ of the total population, survived the entire integration time. 96\% of clones were ejected from the solar system on hyperbolic or parabolic orbits which again points to an origin for Chiron beyond Neptune. Approximately 1\% hit Jupiter, and the remaining 1\% hit the Sun, Saturn, Uranus or Neptune. \par
Using the best-fit line we find that if the decay had remained exponential then 99.99\% of the clones would have been gone by 8.5 Myr ago. We use this time as the upper limit to the time at which Chiron first entered the Centaur region.

\subsection{Close Encounters}
The total number of close encounters between Chiron clones and the giant planets was 24,196,477. 15,130,506 of these occurred while clones were in the Centaur region.\par
During their time in the Centaur region, clones experienced a close encounter on average every 5 kyr. Table~\ref{CE_centaur} shows the number of these close encounters by planet.\par
As expected, clones had the highest numbers of close encounters with Saturn and Uranus, followed by Neptune and then Jupiter. \par
Table~\ref{CE_severity2} lists the percentage of close encounters which occurred in the Centaur region by severity. It can be seen that the lower the severity, the greater the number of close encounters. There were only 48 severe and exactly zero extreme close encounters. These results show that encounters close enough to tidally disrupt Chiron or any ring system around Chiron are extremely rare events.\par
Thus, it is unlikely that any ring structure around Chiron was created by tidal disruption due to a planetary close encounter, and barring ring dispersal by viscous spreading, it is possible that any ring structure around Chiron has survived its journey through the Centaur region and is in fact primordial.

\begin{table} [h]

\caption{Close encounters of Chiron clones with each giant planet while clones were in the Centaur region.}
\begin{tabular} {|c|c|c|}
\hline
Planet&Number\\
\hline
Jupiter&553182\\
Saturn&6978716\\
Uranus&4567440\\
Neptune&3031168\\
\hline
\end{tabular}

\label{CE_centaur}
\end{table}

\begin{table} [h]

\caption{The percentage of close encounters of Chiron clones with the giant planets by severity while clones were in the Centaur region.}
\begin{tabular} {|c|c|}
\hline
Severity&Percent\\
\hline
Very Low&89\\
Low&11\\
Moderate&0.03\\
Severe&0\\
Extreme&0\\
\hline
\end{tabular}

\label{CE_severity2}
\end{table}

\subsection{Dynamical Class of Chiron}
The dynamical classes of 1,246 clones were determined.
Table~\ref{dynamical} shows the percentage of clones in each dynamical class, and the mean Centaur lifetime of clones in each class. 95\% of the sampled clones were classified as random-walk Centaurs, with the remaining 5\% being classified as resonance hopping Centaurs.\par
The difference in mean Centaur lifetime between the two classes is stark. The mean Centaur lifetime for the resonance hopping clones was approximately twice as long as that of random-walk clones.\par
We hypothesise that the large difference is caused by resonance sticking in mean motion resonances of resonance hopping clones having the effect of prolonging their dynamical lifetimes. This is supported by the work of \citet{2009Icar..203..155B}. The top of Figure~\ref{res_hopper_clone_1} shows the behavior of the semi-major axis of the orbit of one of the longest lived resonance hopping clones. In the figure, the semi-major axis spends about 5 Myr oscillating about the 2:3 mean motion resonance of Saturn centered at 12.5 au. Notice the horizontal band feature which covers this period of time. A shorter band centered at 15.1 au is caused by the exterior 1:2 mean motion resonance of Saturn.\par
Examination of other resonance hopping clones also showed relatively long periods of time for which each clone was trapped in one or more mean motion resonances. We conclude that resonance sticking acts to significantly prolong the lives of resonance hopping clones. Other notable resonances entered into by clones include the exterior 3:4, 4:7, and 1:3 resonances of Saturn; the Trojan or 1:1 resonance of Saturn, the interior 3:2 resonance of Uranus; and the interior 3:2 and 4:3 resonances of Neptune.\par
The bottom diagram in Figure~\ref{res_hopper_clone_1} shows the log-log plot used to classify the clone. It can be seen that it only takes one data point with a relatively large residual to cause a clone to be classified as resonance hopping.\par
The top diagram in Figure~\ref{res_hopper_clone_579} shows another example of a resonance hopping clone. In contrast to the clone in Figure~\ref{res_hopper_clone_1} which spends most of its time in one resonance, this clone spends most of its time hopping between mean motion resonances of the giant planets. Two of these resonances were positively identified as the 4:3 and 3:2 mean motion resonances of Neptune by observing the libration of their primary resonance angles.\par
The bottom diagram shows a close up of the time spent in the 4:3 mean motion resonance of Neptune before and after data smoothing. The smoothed data set has a mean semi-major axis value that is only 0.07 au away from the 4:3 mean motion resonance of Neptune, located at 24.89 au.\par
Figure~\ref{res_angle_4_3} shows the primary resonance angle associated with the 4:3 mean motion resonance of Neptune for the clone in Figure~\ref{res_hopper_clone_579} over the same time interval. The angle is defined by $4\lambda_N - 3\lambda - \bar{\omega}$ where $\lambda_N$ is the mean longitude of Neptune. It can be seen that this angle librates. 

\begin{table} [h]

\caption{The percentage of clones and mean Centaur lifetime by dynamical class. Random-walk dominates in quantity, but resonance hopping clones have about twice the mean Centaur lifetime as random-walk clones due to resonance sticking.}
\begin{tabular} {|c|c|c|}
\hline
Class&Percent&Avg. Centaur Life (Myr)\\
\hline
Resonance Hopping&5&1.1\\
Random-Walk&95&0.52\\
\hline
\end{tabular}

\label{dynamical}
\end{table}

\begin{figure*} [h]
\begin{center}
\includegraphics[width = \columnwidth]{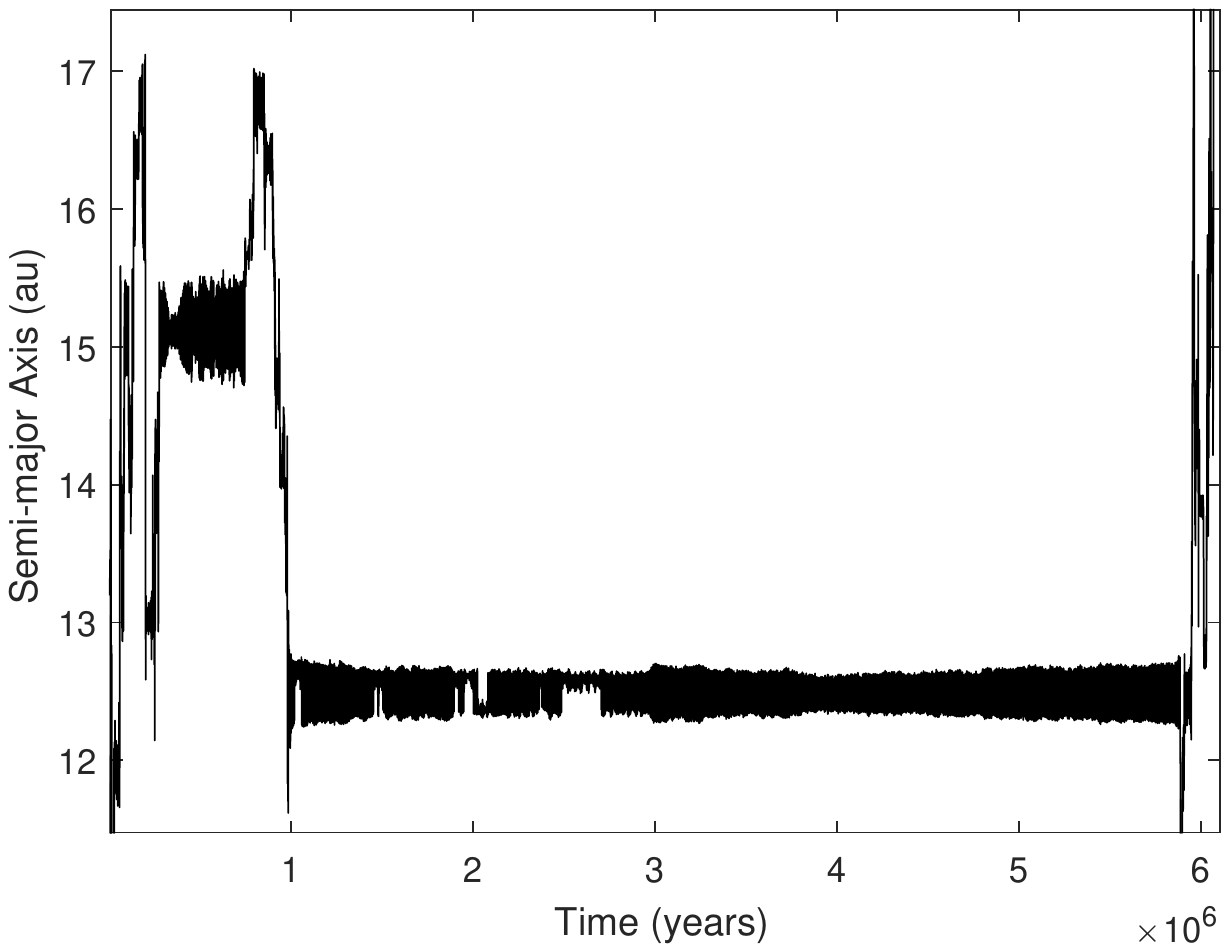}
\includegraphics[width = \columnwidth]{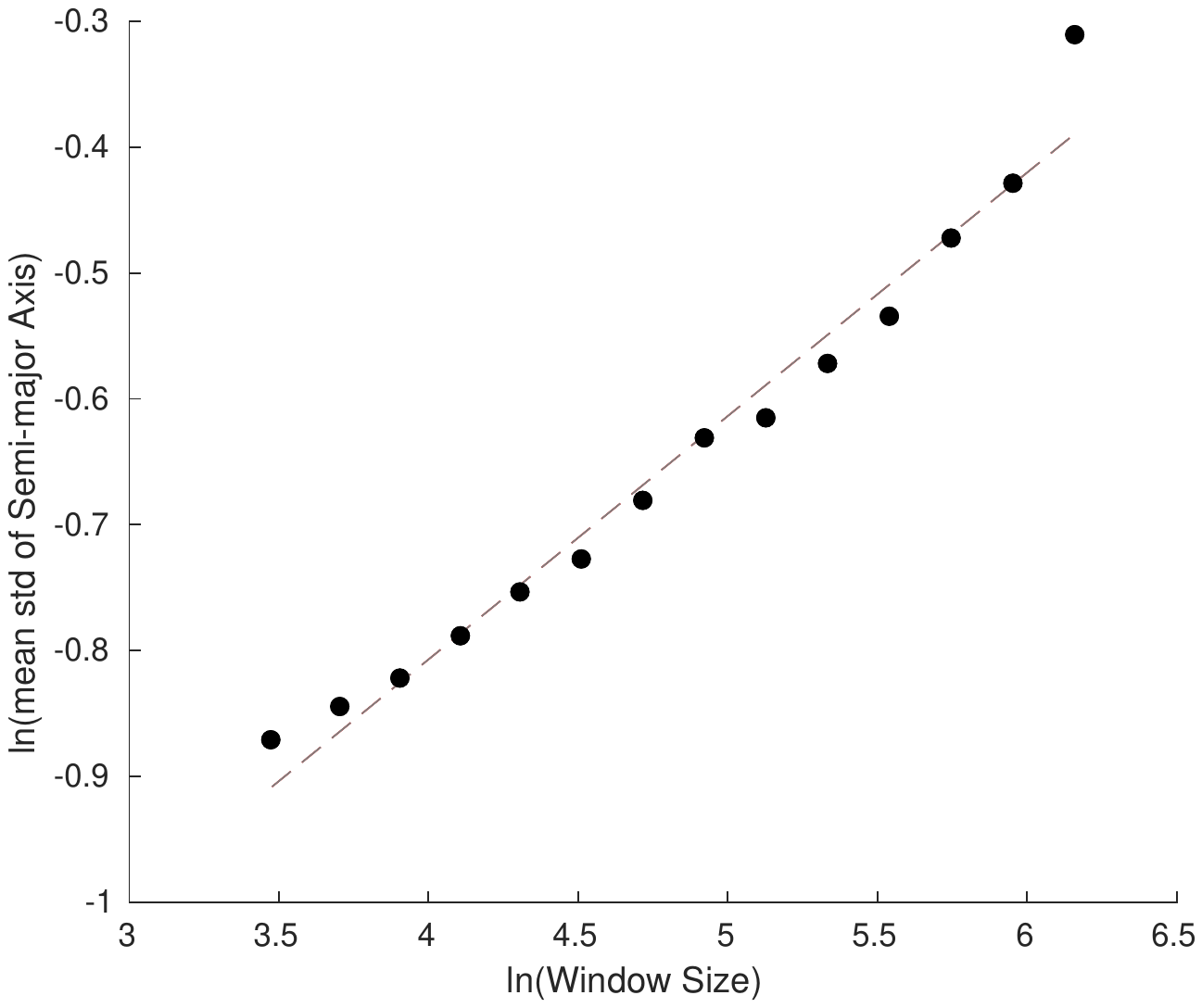}
\caption{Top - an example of a resonance hopping clone. Note the long horizontal band feature. This clone spends about 5 Myr oscillating about the 2:3 mean motion resonance of Saturn located at 12.5 au. A shorter band centered at 15.1 au is caused by the exterior 1:2 mean motion resonance of Saturn. Bottom - the log-log plot used to identify the dynamical class of the clone in the top diagram. Notice the one data point at a larger distance from the trendline than the others. This is characteristic behavior for resonance hopping Centaurs. The Hurst exponent for this clone was 0.193, and its linear regression coefficient was 0.971. The maximum residual was 0.08.}
\end{center}
\label{res_hopper_clone_1}
\end{figure*}


\begin{figure*} [h]
\begin{center}
\includegraphics[]{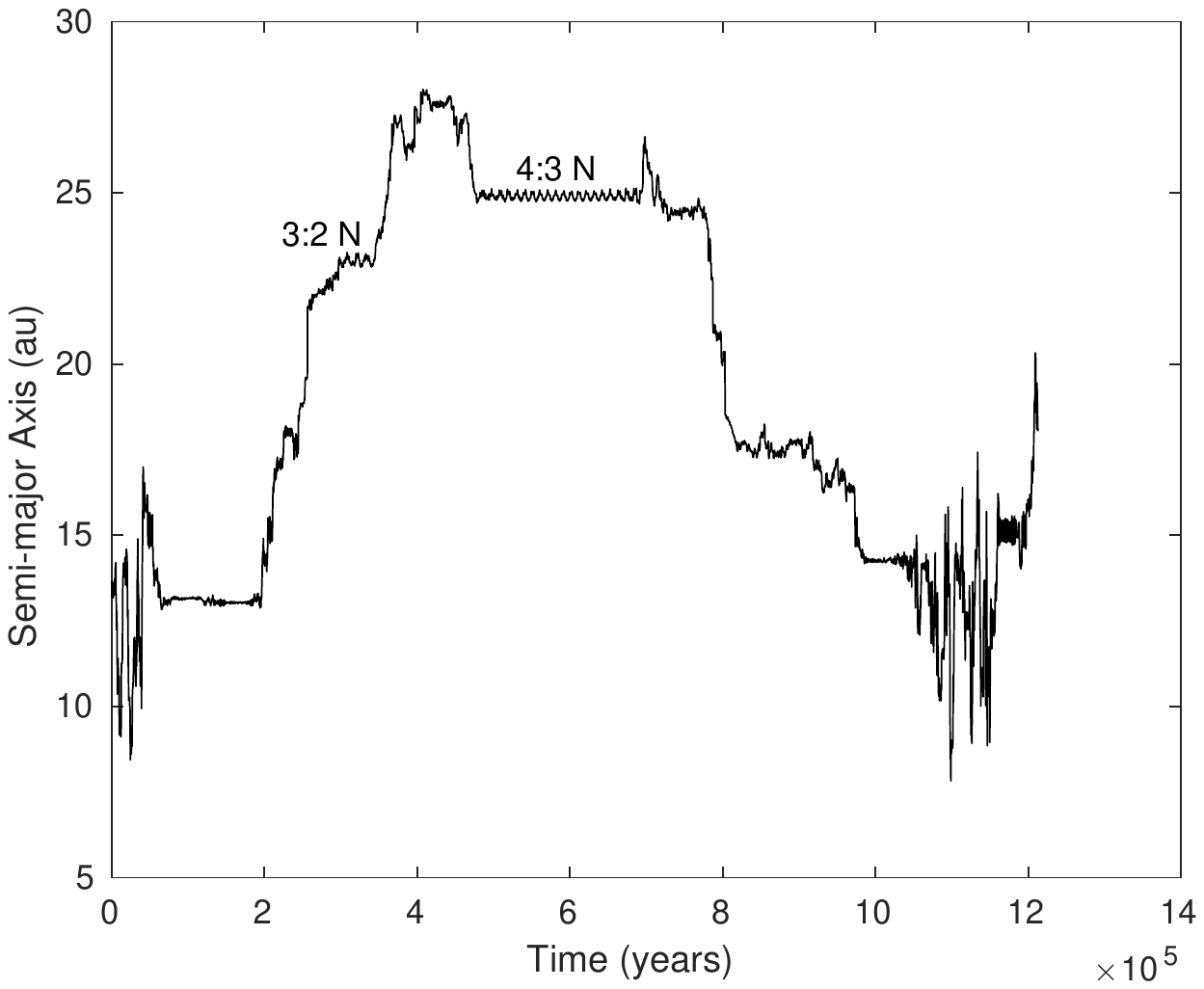}
\includegraphics[]{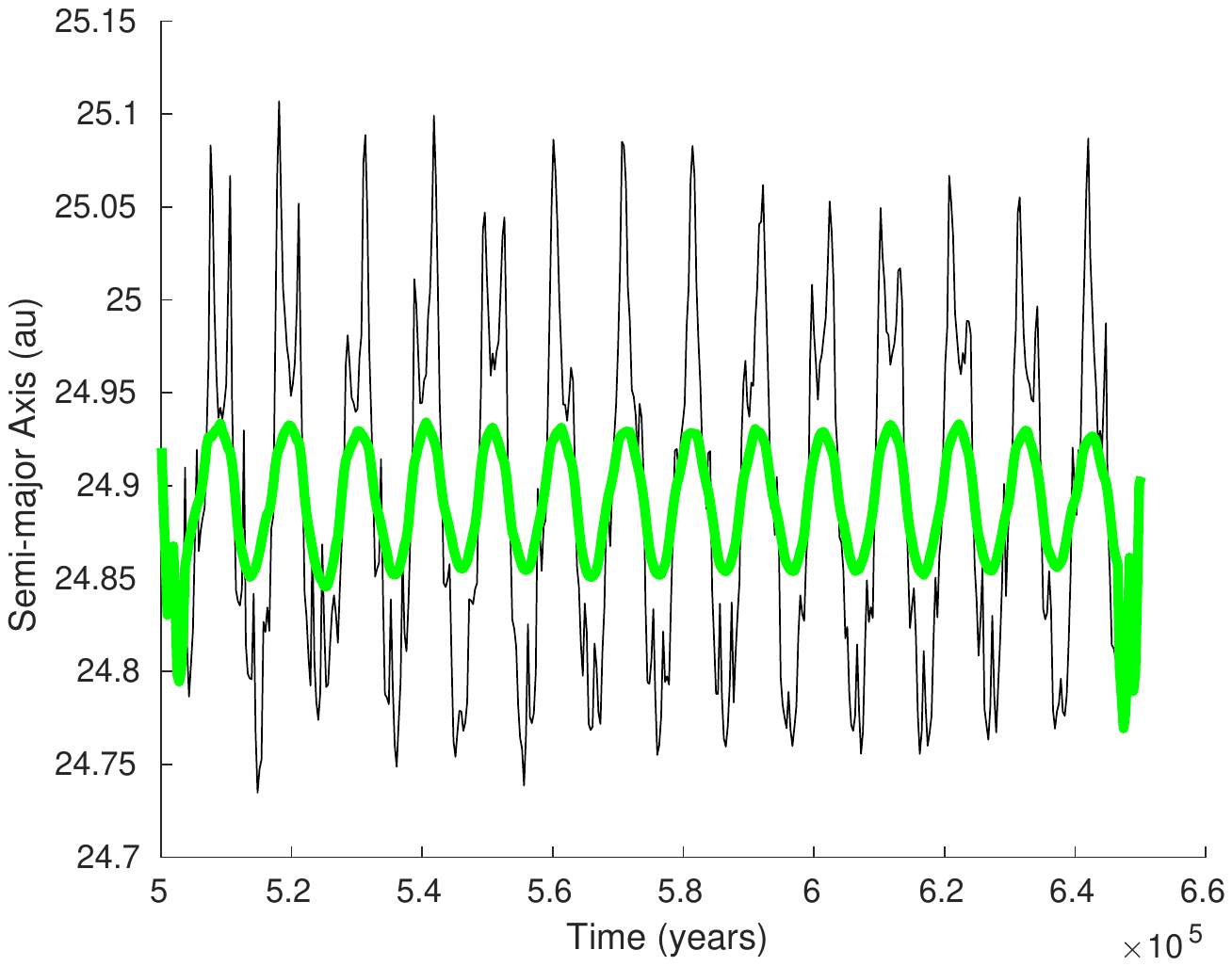}

\caption{Top - another example of a resonance hopping clone. This clone spends most of its time trapped in various mean motion resonances of the giant planets. Two resonances were positively identified as the 4:3 and 3:2 mean motion resonances of Neptune. These are labeled in the figure. The Hurst exponent was 0.534, the linear regression coefficient 0.9937, and the maximum residual was 0.08. Bottom - a close up of the time spent in the 4:3 mean motion resonance of Neptune before and after data smoothing. The mean value of the smoothed data set was 24.89 au which is about 0.07 au away from the 4:3 mean motion resonance of Neptune.}
\end{center}
\label{res_hopper_clone_579}
\end{figure*}

\begin{figure*} [h]
\begin{center}
\includegraphics[]{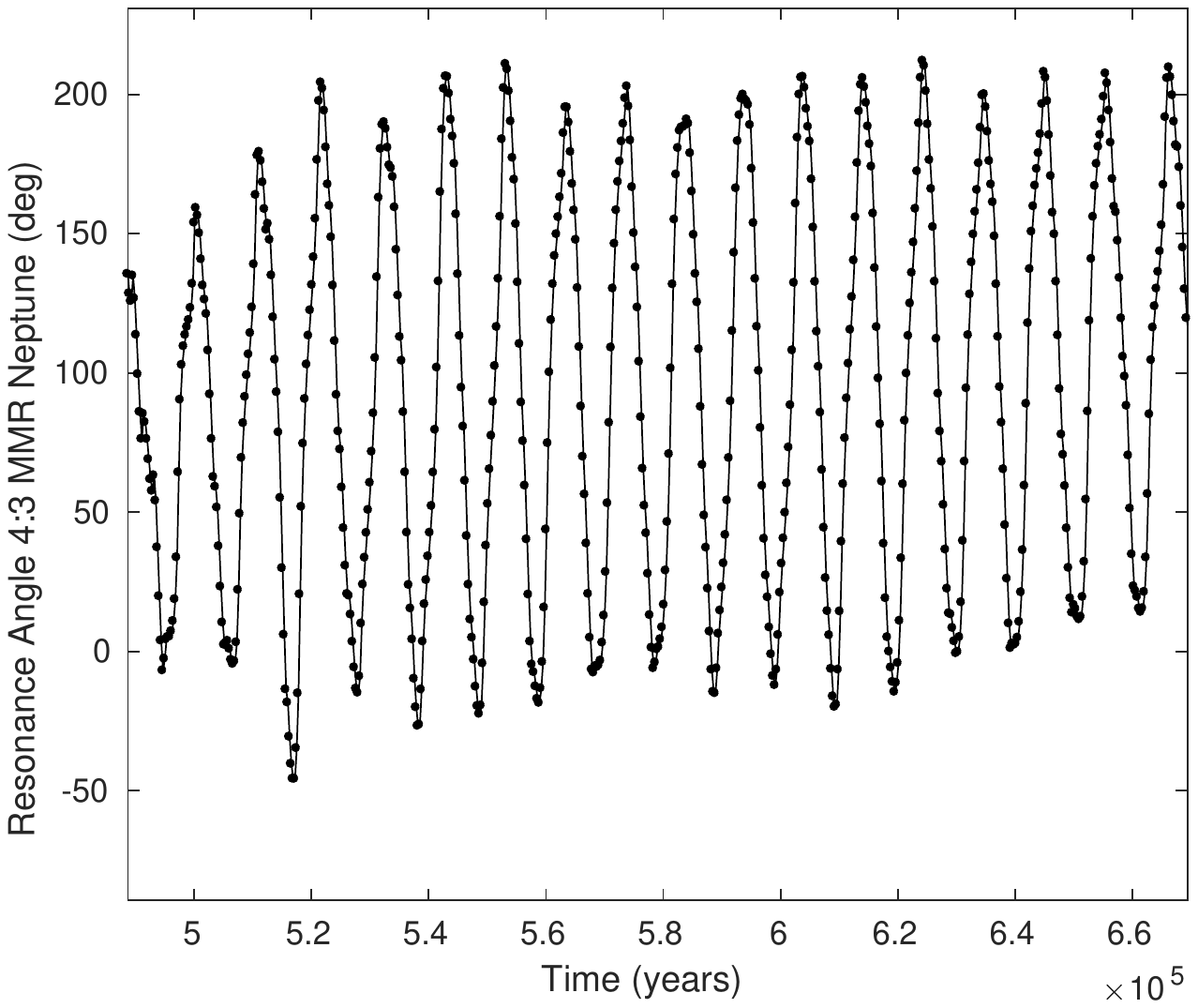}
\caption{The primary resonance angle of the 4:3 mean motion resonance of Neptune defined by $4\lambda_N - 3\lambda - \bar{\omega}$ librates in time.}
\end{center}
\label{res_angle_4_3}
\end{figure*}

Figure~\ref{tp_18018_a_v_t_random_walk} shows an example of a random-walk clone. This clone does not spend the majority of its life trapped in mean motion resonances as can be seen by the lack of long horizontal bands in the figure.

\begin{figure*} [h]
\begin{center}
\includegraphics[width=\columnwidth]{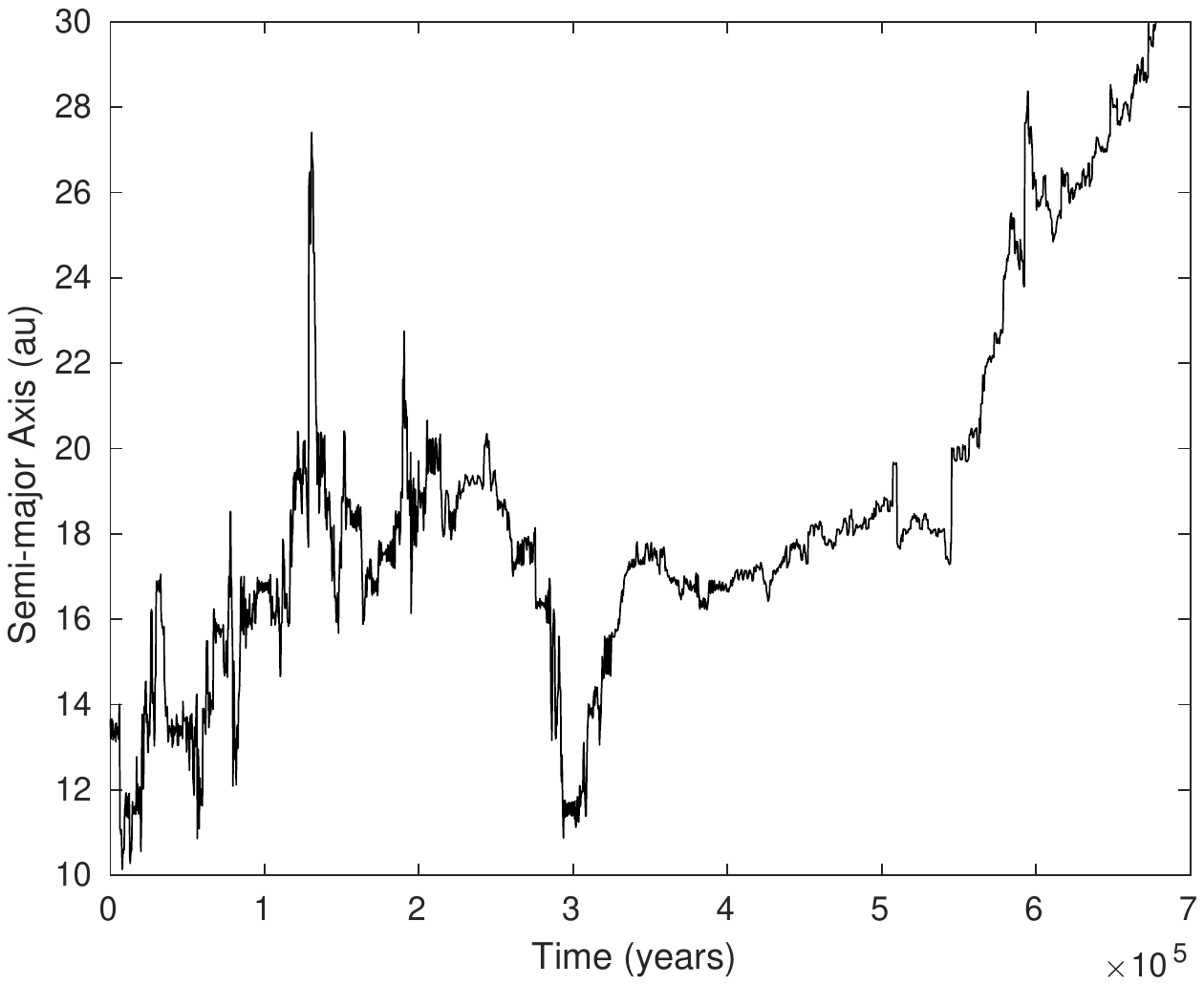}
\includegraphics[width=\columnwidth]{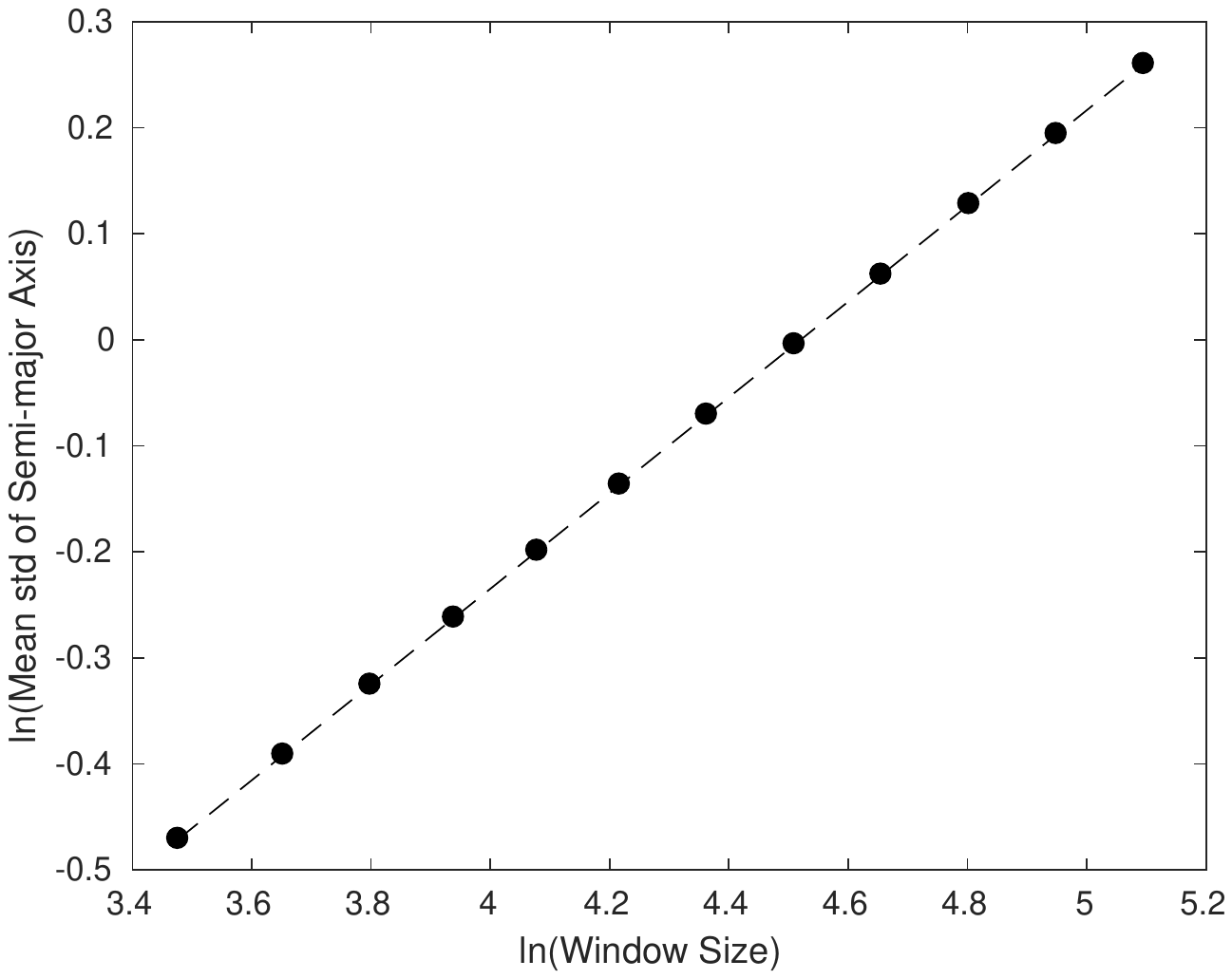}
\caption{Left - An example of a random-walk clone. Notice how the long horizontal bands are absent. Right - the log-log plot used to identify the dynamical class of the clone in the top diagram. Notice the good fit. The linear regression coefficient was 0.9998, and the Hurst exponent for this clone was 0.4514. The maximum residual was 0.008.}
\end{center}
\label{tp_18018_a_v_t_random_walk}
\end{figure*}

The mean Hurst exponent of the random-walk clones is 0.4664 $\pm$ 0.0782 and that of the resonance hopping clones is 0.3572 $\pm$ 0.1530. Here, the error is given by the standard deviation of the mean. It can be seen that Hurst exponents of random-walk clones are more well defined than those of resonance hopping clones as the standard deviation of the mean of the Hurst exponents of random-walk clones is about half that of the resonance hopping clones.\par
Hurst exponents ranged from -0.1764 to 0.6416 for resonance hopping clones and from 0.1446 to 0.7462 for random-walk clones. 0.85 was the lowest regression coefficient for a random-walk clone, and resonance hopping clones had regression coefficients ranging from -0.33 to 0.99.\par
\citet{2009Icar..203..155B} reported that random-walk Centaurs display Hurst exponents in the range 0.22 - 0.95. We found that only five of our random-walk clones had Hurst exponents outside this range - all of them $<0.22$.\par
Qualitative inspection showed that four of these five could be classified as resonance hopping Centaurs as they spent the majority of their lives in mean motion resonances. The fifth clone displayed both random walk and resonance hopping behavior, but spent most of its time experiencing random-walk evolution. The fit of that clone's log-log plot had a regression coefficient of only 0.85, which is more than three standard deviations away from the mean value of 0.9947 $\pm$ 0.0089 for random-walk clones.\par
Furthermore, the outliers also had another thing in common - of the total time spent in resonances, each spent the majority of that time in only one strong resonance and did not jump into any other strong resonances. An example of one of these five outliers is shown in Figure~\ref{tp_18372_a_v_t}.\par
This particular clone spends 66$\%$ of its life in the 2:3 mean motion resonance of Saturn and never jumps to another strong resonance. It was classified as a random-walk clone because its residuals never exceeded 0.0601, but since it spent more time in a resonance than random walking, one could argue that this clone is resonance hopping even though our method classifies it as random-walk. The linear regression coefficient of its log-log plot was 0.88, and its Hurst exponent 0.19.\par
We conclude that our results are in good agreement with those of \citet{2009Icar..203..155B}, but that our technique occasionally misclassifies a clone. A refinement of this technique may be to consider the regression coefficients as well as the residuals as part of the classification procedure.\par
For example, if the regression coefficient of a random-walk clone falls below some critical value, then the clone should be classified manually. That is, classify it using qualitative inspection of the clone's semi-major axis behavior over time. The exact critical value to use we will leave open for now. \par
Another factor to consider is the distance of the Hurst exponent from the mean. All five of the outlying random-walk clones had Hurst exponents more than three standard deviations away from the mean. A refinement of the technique may be to manually classify any clones with outlying Hurst exponents. It remains to be seen if all outliers spend most of their lives in just one strong resonance or if this is just coincidental.\par

\begin{figure*} [h]
\begin{center}
\includegraphics[]{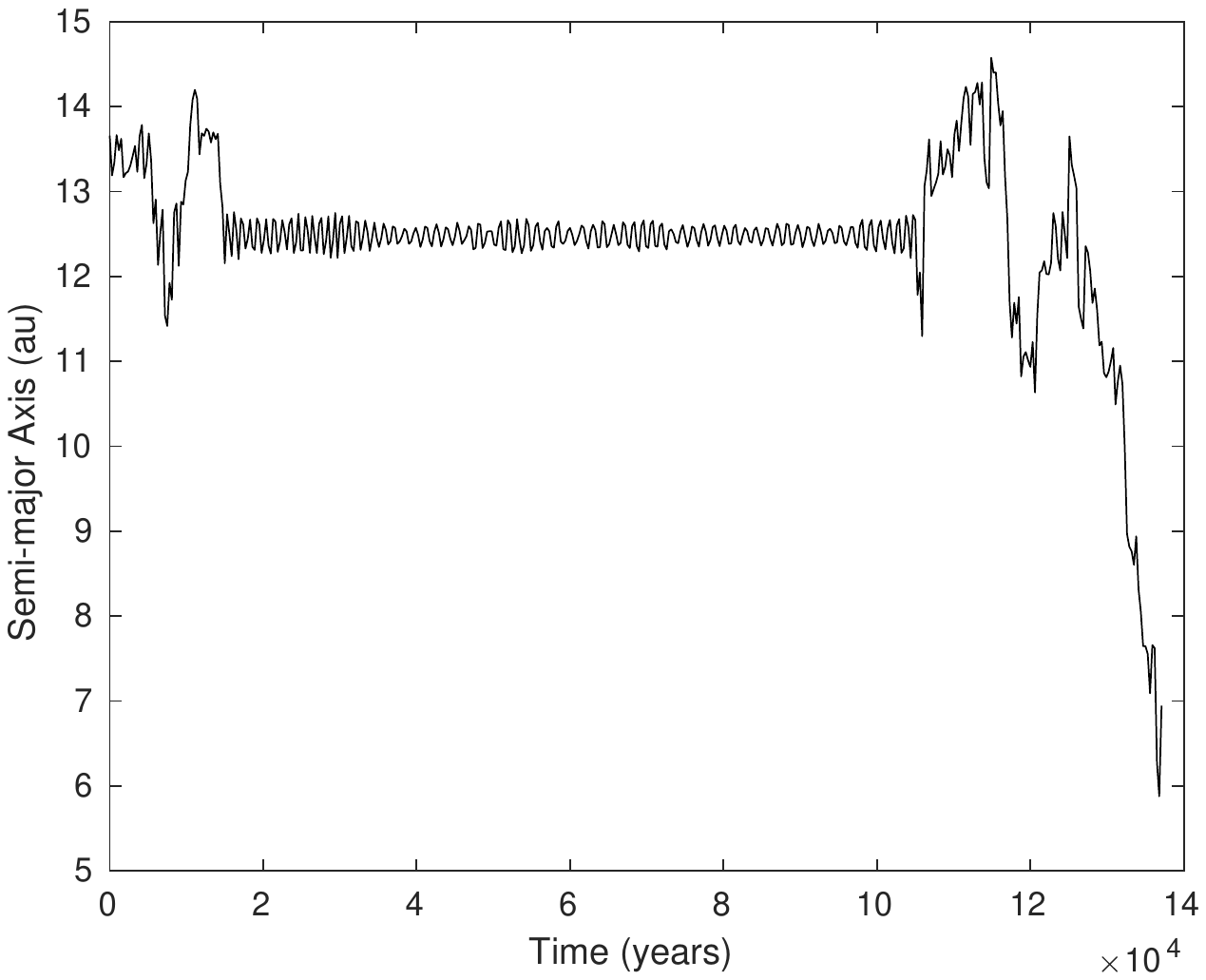}
\caption{A random-walk clone that spent most of its life in the 2:3 mean motion resonance of Saturn located at 12.5 au. Though its residuals were $\le 0.0601$, one could argue that it is a resonance hopping clone.}
\end{center}
\label{tp_18372_a_v_t}
\end{figure*}

\subsection{MEGNO and Lifetime Maps}
Figure~\ref{lifetime_map} shows the chaotic lifetimes of orbits in the region bound by $13 \textnormal{ au} \le a \le 14 \textnormal{ au}$ and $0 \le e \le 0.5$. It can be seen that most orbits with $e\ge 0.23$ have lifetimes typically $\le 0.01$ Myr which are noticeably shorter than the lifetimes of orbits of much lower eccentricity.\par
Chiron, located at the point (13.64 au, 0.38) lies in this region of relatively short lifetimes. Orbits with $a=13\textnormal{ au}$ and eccentricity of 0.23 just begin to cross the orbit of Saturn. All orbits with eccentricities above about 0.28 are Saturn crossing. This allows strong close encounters between objects on those orbits and the giant planet to occur immediately which explains why most orbits with $e \ge 0.28$ have lifetimes $\le 0.01$ Myr - the lowest in the map.\par
One exception to this is the bump-like feature centered at 13.4 au, with a width of about 0.2 au. Orbits within the bump with eccentricities as high as $0.35$ have lifetimes noticeably greater than 0.01 Myr.\par
For example, there are orbits in the bump with $e\ge 0.28$ with lifetimes of 0.1 Myr which is an order of magnitude longer than most other orbits in the map with $e\ge 0.28$. Also of note is a cluster of orbits within the bump near $e=0.1$ for which lifetimes can reach as high as 1 Myr - the longest in the map. We hypothesise that the bump feature is caused by resonance sticking in the 3:5 mean motion resonance of Saturn located at 13.4 au.\par
Small objects which get stuck in this resonance could have their chaotic lifetimes extended in the same way that the Centaur lifetime was extended for a clone stuck in the 2:3 mean motion resonance of Saturn as seen in Figure~\ref{res_hopper_clone_1}. It should be noted, however, that most orbits located at 13.4 au with eccentricities below 0.06 have lifetimes noticeably shorter than 1 Myr.\par
This implies that small objects in this region of phase space are either not being captured in the resonance or are staying in the resonance for shorter times which results in lower lifetimes. This may be caused by the decreasing width of the resonance for smaller eccentricities.\par
Such behavior of resonances has been seen before. For example, \citet{MurrayCD:1999} observed the same behavior for the 3:1 and 5:3 interior mean motion resonances of Jupiter located in the main asteroid belt.\par
Another bump of longer lifetimes which reach as high as 1 Myr is found between 13.9 au and 14 au with $e\le 0.05$. The low eccentricity of orbits in this bump help insulate them from destabilising close encounters with Saturn and Uranus. Though their lifetimes of 1 Myr are relatively long compared to other orbits in the figure, this is still much shorter than the age of the solar system and so these orbits should be viewed as being only relatively stable.\par
Figure~\ref{megno_map} is the MEGNO map of the same region of phase space. Almost the entire region, including the current orbit of Chiron, is highly chaotic. Two features of relatively lower chaos stand out. One island centered around 13.4 au with $0.1 \le e \le 0.15$ and a pair of islands between 13.9 au and 14 au with $e<0.04$. Here, the MEGNO parameter reaches as low as 2.5. Two tinier islands can be seen between 13.7 au and 13.9 au.\par
By comparison of the two maps, it can be seen that these islands are also embedded within regions of relatively long lifetimes which can reach as high as 1 Myr making these islands regions of lower chaos and longer lifetimes.\par
It can also be seen that the two bumps of relatively long lifetimes found in the lifetime map also contain some orbits with lifetimes of 1 Myr which are also highly chaotic. Orbits which are chaotic but have a relatively long lifetime are said to display stable chaos.\par
Chiron, however, cannot be shown to display stable chaos as it has a highly chaotic orbit and relatively short lifetime.

\newpage
\begin{figure*} [h]
\begin{center}
\includegraphics[width=10cm,angle =270 ]{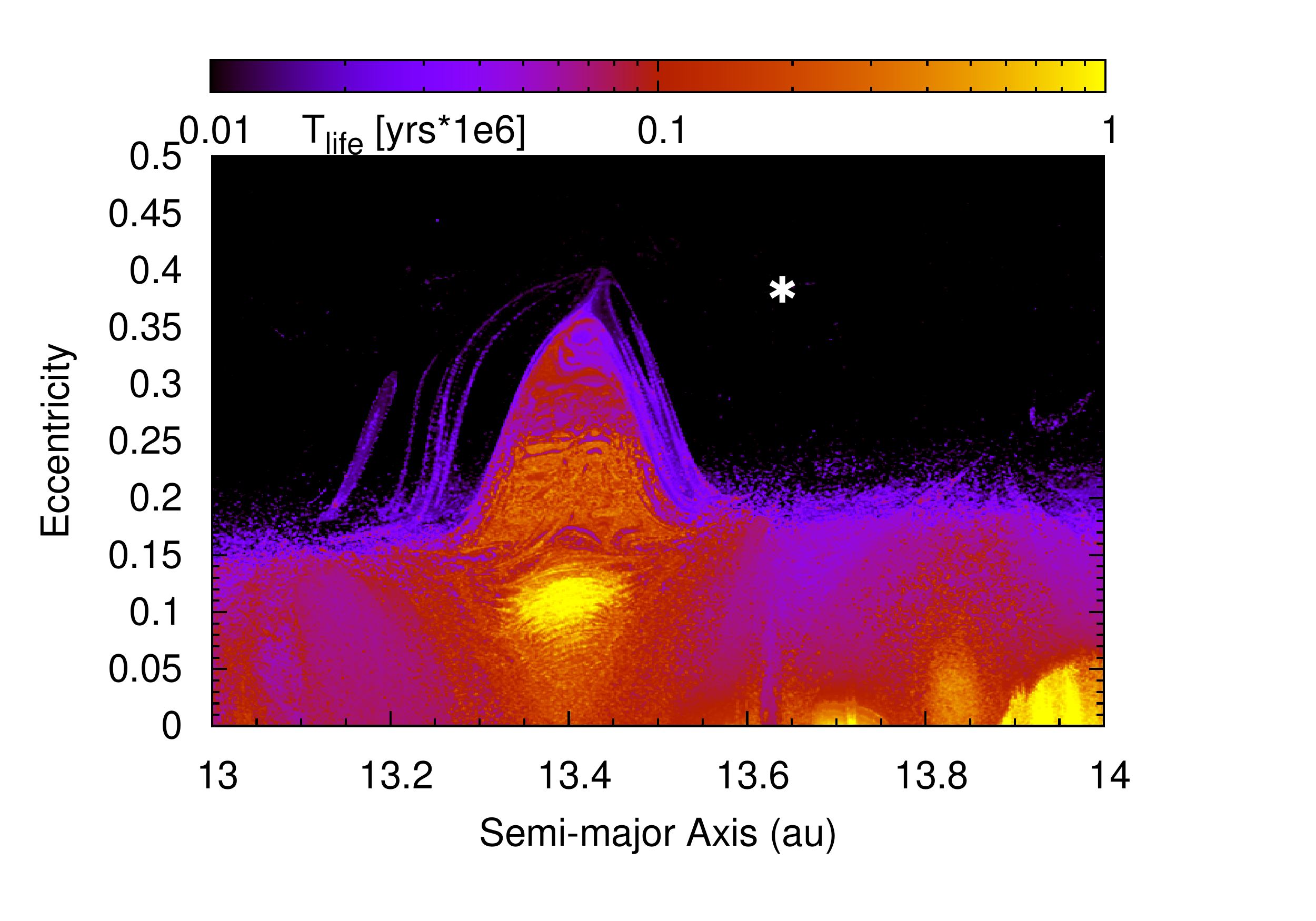}
\caption{The chaotic lifetime map in $a-e$ space. Chaotic lifetime is the time to be removed from the simulation and not dynamical lifetime. However, the dynamical lifetime is greater than or equal to the chaotic lifetime. Chiron is shown as the star at the point (13.64 au, 0.38). A feature which stands out is the bump centered at 13.4 au which has a width of about 0.2 au and a height of about 0.35. We hypothesise that the cause of the bump is resonance sticking in the 3:5 mean motion resonance of Saturn which prolongs the lifetimes of test particles which get trapped in the resonance. A smaller bump can be seen between 13.9 au and 14 au with $e\le 0.05$. There is also a tiny bump in lifetimes up to 1 Myr between 13.7 au and 13.75 au.}
\end{center}
\label{lifetime_map}
\end{figure*}

\newpage
\begin{figure*} [h]
\begin{center}
\includegraphics[width=10cm,angle = 270]{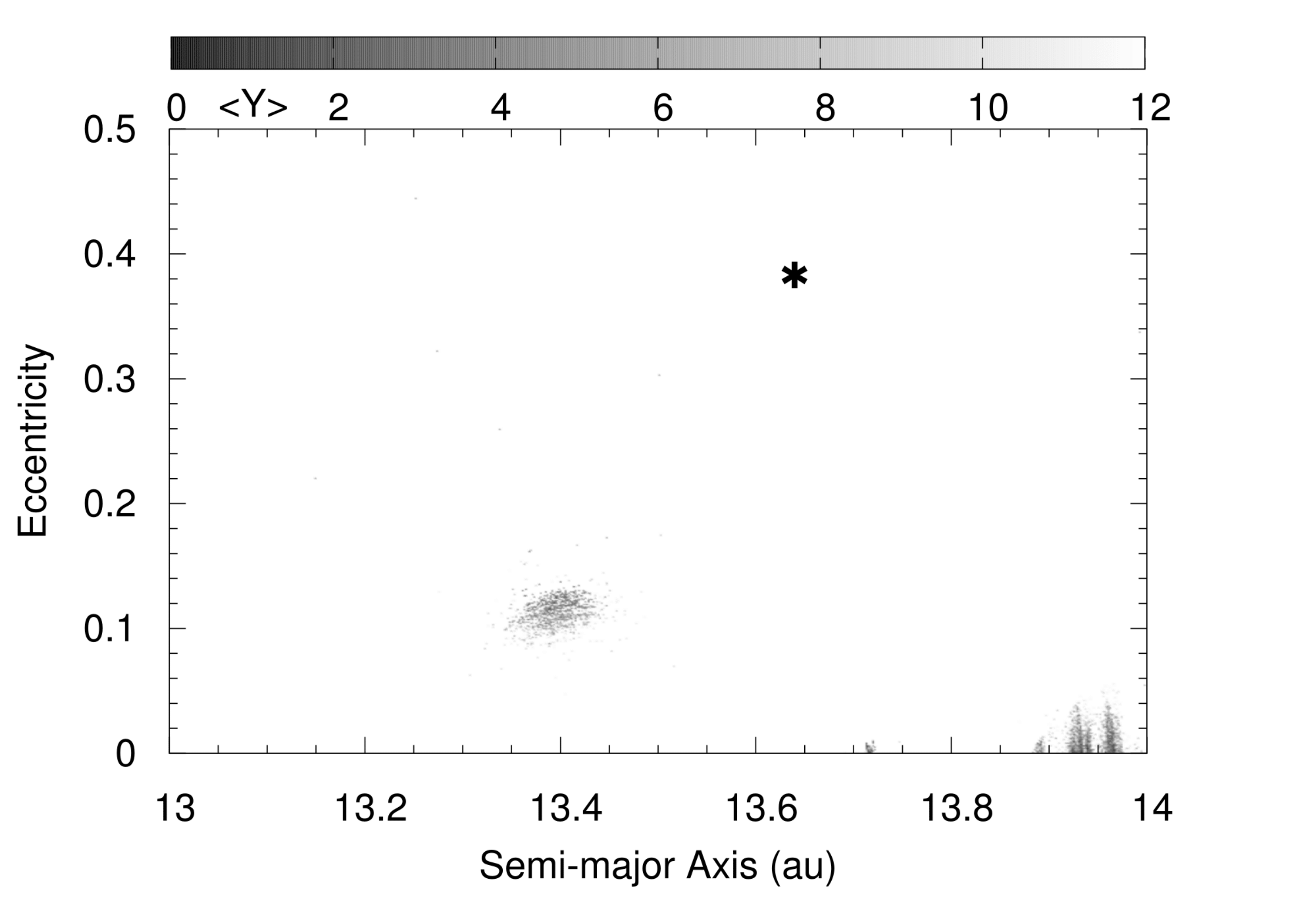}
\caption{The MEGNO map in $a-e$ space. Chiron is shown as the star at the point (13.64 au, 0.38). Nearly the entire region is highly chaotic. There are a few small islands of orbits with relatively low chaos. One is centered near 13.4 au with $0.1 \le e \le 0.15$ where the MEGNO parameter can reach as low as 3.5. Two others can be seen between 13.9 au and 14 au in which the MEGNO parameter reaches as low as 2.5. Two tinier islands can be seen between 13.7 au and 13.9 au.}
\end{center}
\label{megno_map}
\end{figure*}

\newpage
\section{CONCLUSIONS}
Using the technique of numerical integration of nearly 36,000 clones of the Centaur Chiron, we found the backwards half-life of Chiron's orbit to be 0.7 Myr and showed that Chiron likely entered the Centaur region from somewhere beyond Neptune within the last 8.5 Myr.\par
Close encounters between Chiron and the giant planets severe enough to tidally disrupt Chiron or any ring system in a single pass were found to be extremely rare, and thus the origin of any ring structure is unlikely the result of tidal disruption of Chiron due to a planetary close encounter. \par
This led us to conclude that any supposed ring system around Chiron could be primordial barring ring dispersal by viscous spreading. Our results are similar to those of \cite{2017AJ....153..245W} and \citet{AraujoRAN:2016} for the ringed Centaur Chariklo. In those studies, close encounters severe enough to severely damage or destroy the ring structure around Chariklo were also found to be very rare.\par
We also showed that the orbit of Chiron lies in a region of phase space that is both unstable and highly chaotic and that the chaotic lifetime of Chiron is likely to be $\le 0.01$ Myr. Resonance sticking was shown to have the ability to prolong the Centaur lifetime of Chiron clones by up to two orders of magnitude  beyond its chaotic lifetime. Resonance sticking in the 2:3 exterior mean motion resonance of Saturn was cited as a strong example of this.\par
The dynamical classes of a sample of 1,246 clones were determined while these clones were in the Centaur region. It was found that 95\% of clones in the sample were categorized as random-walk Centaurs, and the remaining 5\% categorized as resonance hopping Centaurs. Because of resonance sticking, the mean Centaur lifetime of resonance hopping clones was about twice that of random-walk clones.\par 
MEGNO and lifetime maps were made of the region in phase space bound by $13 \textnormal{ au}\le a \le 14\textnormal{ au}$ and $e\le 0.5$ which included the orbit of Chiron. It was found that nearly the entire region is highly chaotic with relatively small islands of lower chaos. Other small islands of stable chaos (high chaos and relatively long lifetime) were found.\par
Most orbits with eccentricities $\ge 0.28$ had the lowest chaotic lifetimes in the map of $\le 0.01$ Myr due to the crossing of Saturn's orbit. However, some test particles in orbits with $e\ge 0.28$ and semi-major axes within about 0.1 au of the exterior 3:5 mean motion resonance of Saturn located at 13.4 au were shown to have lifetimes up to 0.1 Myr even for orbits with eccentricities up to about 0.35.\par
More research is needed to determine conclusively if the structure around Chiron is a ring system. It is not known if rings around small bodies are rare or commonplace. If future discoveries reveal that ringed Centaurs are common, it would suggest a common mechanism for the creation of the rings. \par
If on the other hand ringed Centaurs are found to be rare then this would suggest a more serendipitous origin for rings. The authors encourage more searches for rings around other small bodies to help answer this question.

\section{Acknowledgments}
Thanks to the referee for their input on this work. Part of the numerical simulations were performed using a high performance computing cluster at the Korea Astronomy and Space Science Institute. TCH acknowledges KASI grant \#2016-1-832-01 and \#2017-1-830-03.
\end{document}